\documentclass[preprint,aps]{revtex4}

\usepackage{graphicx}

\def\beq{\begin{equation}}
\def\eeq{\end{equation}}
\def\bea{\begin{eqnarray}}
\def\eea{\end{eqnarray}}
\def\Im{{\rm Im}}

\def\nn{\nonumber}
\def\sss{\scriptscriptstyle}

\def\barp{{\raise.35ex\hbox
{${\sss (}$}}---{\raise.35ex\hbox{${\sss )}$}}}
\def\bdbarp{\hbox{$B_d$\kern-1.4em\raise1.4ex\hbox{\barp}}}
\def\bsbarp{\hbox{$B_s$\kern-1.4em\raise1.4ex\hbox{\barp}}}
\def\ks{K_{\sss S}}

\def\roughly#1{\mathrel{\raise.3ex\hbox
{$#1$\kern-.75em\lower1ex\hbox{$\sim$}}}}

\def\even{{\sss \rm even}}

\def\adir00{{a_{\sss dir}^{00}}}

\def\B00{B^{00}}
\def\Bp0{B^{+0}}

\def\dsp{\displaystyle}

\def\prd#1#2#3{{ Phys.\ Rev.} {\bf D#1}, #3 (19#2)}

\def\prl#1#2#3{{ Phys.\ Rev.\ Lett.} {\bf #1}, #3 (19#2)}

\begin{document}

\preprint{OITS-718}
\preprint{IMSc-2002/07/18}

\title{Weak Phase $\gamma$ Using Isospin Analysis and Time Dependent Asymmetry
  in $B_d\to \ks\pi^+\pi^-$}
\author{N. G. Deshpande}
\email{desh@oregon.uoregon.edu}
\affiliation{Institute of Theoretical Science, University of Oregon,
  Eugene, OR94703, USA} 
\author{Nita Sinha}\email{nita@imsc.ernet.in} 
\author{Rahul Sinha}\email{sinha@imsc.ernet.in}
\affiliation{The Institute of Mathematical Sciences, C.I.T Campus, Taramani, 
Chennai 600113, India.}
\date{\today}
\begin{abstract}            
  We present a method for measuring the weak phase $\gamma$ using
  isospin analysis of three body $B$ decays into $K\pi\pi$ channels.
  Differential decay widths and time dependent asymmetry in $B_d\to
  \ks\pi^+\pi^-$ mode needs to be measured into even isospin $\pi\pi$
  states. The method can be used to extract $\gamma$, as well as, the
  size of the electroweak penguin contributions. The technique is free
  from assumptions like SU(3) or neglect of any contributions to the
  decay amplitudes. By studying different regions of the Dalitz plot,
  it is possible to reduce the ambiguity in the value of $\gamma$.

\end{abstract}

\maketitle

Time dependent measurements of asymmetries of decay modes of $B_d$
into $CP$ eigenstates \cite{CarterSanda} are very useful in
determining angles of the Unitarity Triangle.  This technique is
particularly significant, since weak phases can be extracted without
any theoretical uncertainty from modes whose amplitudes have a single
weak phase. The time dependent $CP$ asymmetry in the golden mode $B_d\to
J/\!\psi K_{\sss S}$, thus yields information on $\sin 2\beta$.  This
method has proved successful in the measurement of $\sin 2\beta =
0.78\pm 0.08$ \cite{betameas}, which is in good agreement
\cite{Hoecker} with theoretical estimates.

Measurement of other angles using modes like $B_d \to\pi^+\pi^-$, are
beset with theoretical uncertainties because the amplitude gets
contributions from tree and penguin diagrams which have different
dependence on weak phases. Nevertheless, the theory error can in
principle be removed using the method of Gronau and London
(GL)~\cite{GL}. This method relies on the assumption of isospin
invariance and the fact that the amplitude for $B^+\to \pi^+\pi^0$
gets contributions only from the tree diagrams (barring a small
contribution from the electroweak penguin). This method will lead to a
measurement of $\sin 2\alpha$.

It is widely believed that $\gamma$ cannot be measured using the time
dependent techniques developed to measure the  phases $\beta$
and $\alpha$. As an alternative, several other methods have been
developed~\cite{gamma-tech} to measure this weak phase.  While
$\gamma$ can be measured cleanly using some of these techniques at a
later date, techniques~\cite{SU3} assuming flavor SU(3), are expected
to provide the first estimates of angle $\gamma$.

In this letter, we propose a method, which uses time dependent
asymmetry in the three body $K\pi\pi$ decay mode of the $B_d$. Our
technique is on almost as good a footing as the Gronau-London method,
and relies on construction of triangles based on isospin analysis. The
extra ingredient that we use is that the tree and the electroweak
penguin pieces of the weak Hamiltonian responsible for $\Delta I=1$
transition have the same strong phase because of the operator
structure of the interaction in the standard model.  However, the
method is free from approximations such as SU(3) symmetry, neglect of
annihilation or rescattering contributions. Further, our method is
sensitive to the relative weak phase between the tree and penguin
contribution, and as such will probe new physics. Recently, several
three body non charmed decay modes of $B$ meson have been observed. In
particular the branching ratios of the modes $B^0\to K^0 \pi^+\pi^-$
and $B^0\to K^+\pi^-\pi^0$ have been measured \cite{CLEO,BELLE} to be
around $5\cdot 10^{-5}$. In fact, even with limited statistics, a
Dalitz plot analysis has been performed and quasi two body final
states have been identified.

The three body decay modes such as $B\to K\pi\pi$ provide valuable
information that can pin down the phases in the standard model.
Importance of these modes was first pointed out by Lipkin, Nir, Quinn
and Snyder (LNQS)\cite{LNQS}, however, their analysis did not
incorporate the large electroweak penguin effects known to be present
in these decays\cite{DH}. These decays are described by six independent isospin
amplitudes $A(I_t,I_{\pi\pi},I_f)$, where $I_t$ stands for the
transition isospin, and describes the transformation of the weak
Hamiltonian under isospin and can take only the values 0 and 1 in the
standard model; $I_{\pi\pi}$ is the isospin of the pion pair and takes
the value 0, 1, and 2 and $I_f$ is the final isospin and can take the
values $1/2$ and $3/2$. Even values of $I_{\pi\pi}$ has the pair of
pions in a symmetric state, and thus have even angular momenta.
Similarly states with $I_{\pi\pi}$ odd must be odd under the exchange
of two pions.  A separation between $I_{\pi\pi}={\rm even}$ and
$I_{\pi\pi}={\rm odd}$ should be possible through a study of the
Dalitz plot.

We shall consider only the $I_{\pi\pi}=0$ and 2 channels in this
paper, and these are described by the three amplitudes
$A(0,0,\frac{1}{2})$, $A(1,0,\frac{1}{2})$, and $A(1,2,\frac{3}{2})$.
The amplitudes for the various decay modes with $I_{\pi\pi}={\rm
  even}$, obey useful isospin relations.  It is straightforward to
derive the following \cite{LNQS}:
\begin{eqnarray}
A(B^+\to K^0(\pi^+\pi^0)_\even)&=& X \nn\\
A(B^0\to K^+(\pi^-\pi^0)_\even)&=& -X \nn\\
A(B^+\to K^+(\pi^+\pi^-)_\even)&=&-\frac{1}{3}X-Y+Z \nn\\
A(B^0\to K^0(\pi^+\pi^-)_\even)&=&\frac{1}{3}X+Y+Z \nn\\
A(B^+\to K^+(\pi^0\pi^0)_\even)&=&-\frac{2}{3}X+Y-Z \nn\\
A(B^0\to K^0(\pi^0\pi^0)_\even)&=&\frac{2}{3}X-Y-Z ~,
\label{iso-relations}
\end{eqnarray}
where 
\begin{eqnarray}
 X=\sqrt{\frac{2}{5}} A(1,2,\frac{3}{2})~,~~~ Y=\frac{1}{3}
A(1,0,\frac{1}{2})~,~{\rm and}~~~ Z=\sqrt{\frac{1}{3}} A(0,0,\frac{1}{2})~.
\nn
\end{eqnarray}
The subscript even represents the isospin of the $\pi\pi$ system. It
is easy to see that Eq.~(\ref{iso-relations}) implies the following
two isospin triangles relations:
\begin{eqnarray}
A(B^+ \to K^0(\pi^+\pi^0)_\even)&=&A(B^0 \to
K^0(\pi^+\pi^-)_\even)+A(B^0\to K^0(\pi^0\pi^0)_\even)~,
  \label{neutral-triangle}\\
A(B^0 \to K^+(\pi^-\pi^0)_\even)&=&A(B^+ \to
K^+(\pi^+\pi^-)_\even)+A(B^+ \to K^+(\pi^0\pi^0)_\even)~,
  \label{charged-triangle}
\end{eqnarray}
and also implies the relation,
\begin{equation}
A(B^+\to K^0(\pi^+\pi^0)_\even)=-A(B^0\to K^+(\pi^-\pi^0)_\even).
\label{iso-rel}
\end{equation}
Decays corresponding to conjugate processes will obey similar
relations.  The iso-triangle represented by
Eq.~(\ref{neutral-triangle}) and its conjugate are the ones that
interest us.

 The decay $B(p_B)\to K(k)\pi(p_1)\pi(p_2)$, (where
$p_B$, $k$, $p_1$ and $p_2$ are the four momentum of the $B$, $K$,
$\pi_1$ and $\pi_2$ respectively) may be described in terms of the
usual Mandelstam variables $\mathrm{s}=(p_1+p_2)^2$, $\mathrm{t}=(k+p_1)^2$ and
$\mathrm{u}=(k+p_2)^2$.  States with $I_{\pi\pi}={\rm even}$ must be symmetric
under the exchange $\mathrm{t}\leftrightarrow \mathrm{u}$. In what follows we shall be
concerned with differential decay rates $d^2\Gamma/(d\mathrm{t} d\mathrm{u})$. These can
be extracted from the Dalitz plot of the three body decays. In
particular, if states with $\mathrm{t}=\mathrm{u}$ are selected, they will be
automatically symmetric in $\pi\pi$. A detailed angular analysis will
of course permit extraction of even isospin $\pi\pi$ events and offer
a larger sample.  Note that $B\to K_s\pi^0\pi^0$ mode being symmetric
in pions, always has pions in isospin even state.

For simplicity we define the amplitudes $A^{+-}$, $A^{00}$ and
$A^{+0}$ corresponding to the modes $B\to K_{\sss S}(\pi^+\pi^-)_{\sss
\rm even}$, $B \to K_{\sss S}(\pi^0\pi^0)_{\sss \rm even}$, and $B \to
K_{\sss S}(\pi^+\pi^0)_{\sss \rm even}$ respectively. It may be
understood that all observables, amplitudes and strong phases depend
on the the two independent Mandelstam variables $\mathrm{t}$ and $\mathrm{u}$, even
though we suppress explicitly stating the `$\mathrm{t}$' and `$\mathrm{u}$' dependence.
These amplitudes may be expressed as follows, purely based on CKM \cite{CKM}
contributions:
\begin{eqnarray}
 \label{eq:amps}
A^{+-}&=& a^{+-} e^{i\delta^{+-}_a}e^{i\gamma}
      + b^{+-} e^{i\delta^{+-}_b}\\
 \label{eq:amps2}
A^{00}&=& a^{00} e^{i\delta^{00}_a}e^{i\gamma}
      + b^{00} e^{i\delta^{00}_b}\\
A^{+0}&=& a^{+0} e^{i\delta^{+0}_a}e^{i\gamma}
      + b^{+0} e^{i\delta^{+0}_b}
\end{eqnarray}
The magnitudes $a^{+-}$, $b^{+-}$, $a^{00}$, $b^{00}$, $a^{+0}$ and
$b^{+0}$ actually contain all possible contributions (tree,
color-suppressed, annihilation, W-exchange, penguin,
penguin-annihilation and electroweak-penguin amplitudes) and include
the magnitudes of the CKM elements.  Their explicit composition is
irrelevant for this analysis, except for the fact that the isospin 3/2
amplitude $A^{+0}$ cannot get contributions from gluonic penguins.
The amplitudes $\overline{A}^{+-}$, $\overline{A}^{00}$,
$\overline{A}^{+0}$, corresponding to the conjugate process
$\bar{B}\to \bar{K}\pi\pi$ can be written similarly with the weak
phase $\gamma$ replaced by $-\gamma$.  In the presence of two
contributions to the amplitude as described in Eq.~(\ref {eq:amps})
and (\ref {eq:amps2}), the direct asymmetry is non-vanishing. The time
dependent $CP$ asymmetry for $B^0(t)\to f$ then has the form,
\begin{eqnarray}
A_{CP}^f(t) &=& \frac{\Gamma({\bar B}^0(t)\to
  f)-\Gamma({B}^0(t) \to f)}{\Gamma({\bar B}^0(t)\to f)+
  \Gamma({B}^0(t) \to f)} ~,\nn\\
&=& a_{\sss \rm dir}^f \cos(\Delta m t)
      +\frac{2\,\Im(\lambda^f)}{1+|\lambda^f|^2} \sin(\Delta m t)~,
\end{eqnarray}
where 
\begin{equation}
a_{\sss \rm dir}^f =\frac{\dsp|\bar{A}^f|^2-|A^f|^2}
   {\dsp|\bar{A}^f|^2+|A^f|^2}~,~~~\lambda_f=\frac{q}{p}\frac{\bar{A}^f}{A^f}~~~{\rm and}~~~ \frac{q}{p}=e^{-2i\beta}~.
\end{equation}

\begin{figure}[htb]
\includegraphics{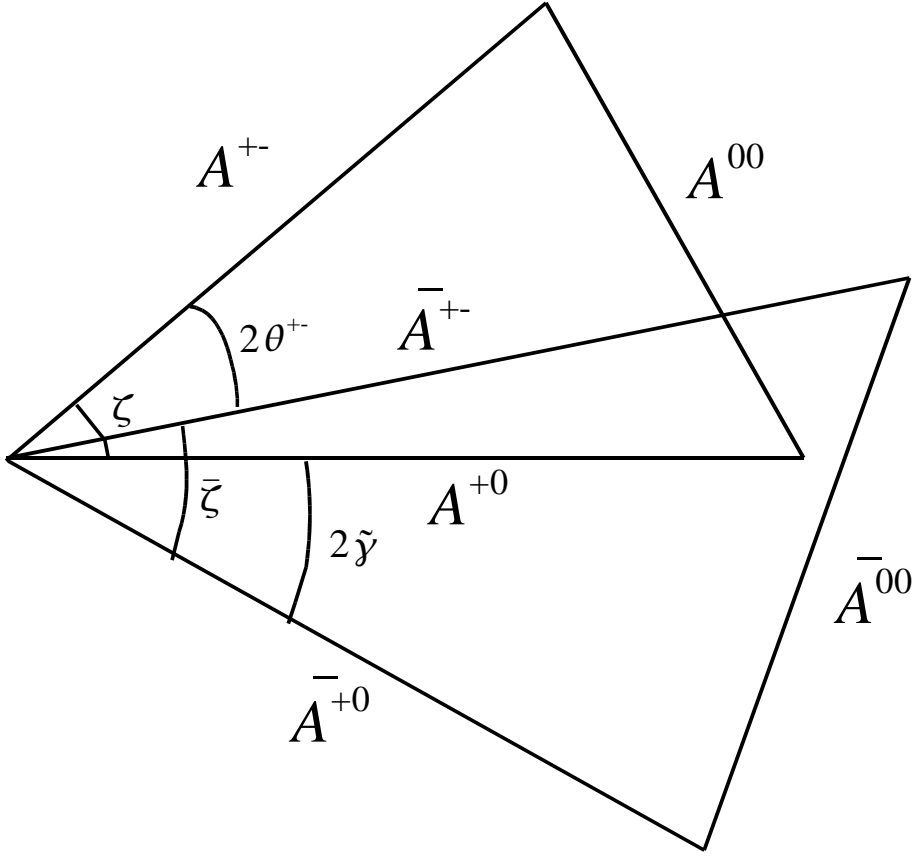}
\caption{\label{fig1}The isospin triangles formed by the $B\to K\pi\pi$
  amplitudes, as represented in Eq.~(\ref{neutral-triangle}) and that
  for the corresponding conjugate processes. Only one orientation of
  the conjugate triangle is depicted, this triangle could have been
  flipped around the base ${\bar A}^{+0}$.}

\end{figure}

Fig.~\ref{fig1} depicts the two triangles formed by the amplitudes
$A^{+-}$, $A^{00}$ and $A^{+0}$ and the corresponding conjugate
amplitudes in isospin space, along with the relative orientations and
defines the angles used in the derivations.  The relative phase
between ${A}^{+-}$ and $\overline{A}^{+-}$ ({\it i.e.}
$\arg(({A}^{+-})^*\overline{A}^{+-})$), defined as $2\theta^{+-}$, can
be obtained from the coefficient of the $\sin(\Delta m t)$ piece in
the time dependent $CP$ asymmetry for the mode $B\to K_{\sss
  S}(\pi^+\pi^-)_{\sss \rm even}$:
\begin{equation}
\frac{2\,\Im(\lambda^{+-})}{1+|\lambda^{+-}|^2}=y^{+-}\,\sin(2\theta^{+-}-2\beta)~.
\end{equation}
where $y^f$ is defined as $y^f=\sqrt{1-(a_{\sss \rm dir}^f)^2}.$ Note
that this measurement involves time dependent asymmetry in the partial
decay rate $d^2\Gamma^{+-}/d\mathrm{t}d\mathrm{u}$ at a fixed
$\mathrm{t}$ and $\mathrm{u}$. Again, events symmetric in $\mathrm{t}
\leftrightarrow \mathrm{u}$ need to be selected in the Dalitz plot.

With the knowledge of $\beta$, the angle $2\theta^{+-}$ may be
regarded as an observable. In addition, measurement of six partial
decay rates $d^2\Gamma^{+0}/d\mathrm{t}d\mathrm{u}$,
$d^2\Gamma^{+-}/d\mathrm{t}d\mathrm{u}$ and
$d^2\Gamma^{00}/d\mathrm{t}d\mathrm{u}$ as well as their conjugates at
the same $\mathrm{t}$ and $\mathrm{u}$ as used for $\theta^{+-}$
determination, now allows us to construct the two triangles in
Fig.~\ref{fig1} with two fold ambiguity. From the figure, we see that
the angle $2{\tilde\gamma}$ between $A^{+0}$ and $\overline{A}^{+0}$
is related to $2\theta^{+-}$ as,
\begin{equation}
\zeta\pm \bar{\zeta} +2{\tilde\gamma}=2 \theta^{+-}~.
\end{equation}
The `plus--minus' sign ambiguity in the above reflects the possibility
of same--side or opposite--side orientation of the triangles. Once
$2\tilde{\gamma}$ is known, it is possible to determine $\gamma$. The
crucial additional information necessary is the observation of Neubert
and Rosner\cite{NR} that the electroweak penguin operators $Q_9$ and
$Q_{10}$ are Fierz-equivalent to the operators $Q_1$ and $Q_2$. The
isospin $3/2$ amplitude $A^{+0}$ is symmetric in the two pions
$(\pi^+\pi^0)$.  Hence, within the Standard Model (SM) only the
operator $(Q_1+Q_2)$ with coefficient $\frac{1}{2}[\lambda_u
(C_1+C_2)-\frac{3}{2}\lambda_t (C_9+C_{10})]$ contributes, while the
operator $(Q_1-Q_2)$ does not.  The amplitude $A^{+0}$ thus has a
common strong phase $\delta=\delta^{+0}_a=\delta^{+0}_b$ arising from
the same quark operator. This phase $\delta$ may be set equal to zero
by convention.

Thus the amplitudes $A^{+0}$ and $A^{-0}\equiv \overline{A}^{+0}$ may
be written as \cite{GPY},
\begin{eqnarray}
A^{+0}&=&(e^{i\gamma}-\delta_{EW}) a^{+0}~,\nn\\
A^{-0}&=& (e^{-i\gamma}-\delta_{EW}) a^{+0}~.
\end{eqnarray}
Here,
\begin{eqnarray}
\delta_{EW}=\frac{-b^{+0}}{a^{+0}}\simeq -\frac{3}{2}
      \left|\frac{\lambda_t}{\lambda_u}\right|
      \frac{C_9+C_{10}}{C_1+C_2}=0.66\pm0.15~,
\label{dew}
\end{eqnarray}
where $\lambda_{q}=V^*_{qb}V_{qs}$.  The angle $2\tilde{\gamma}$ is
then given by
\begin{equation}
 \label{eq:xi-gamma-rel}
\tan\tilde{\gamma} = \frac{\sin\gamma}{\cos\gamma-\delta_{EW}}~.
\label{GP}
\end{equation}
Since the angle $\tilde{\gamma}$ is determined, it follows that angle
$\gamma$ is now calculable from Eq.~(\ref{GP}).

It turns out that we can determine $\gamma$ without having to use the
theoretically computed value of $\delta_{EW}$, given by
Eq.~(\ref{dew}). As we will show below, $\gamma$ can be determined
cleanly by relying only on the Neubert-Rosner observation that the
amplitude $A^{+0}$ has a single common strong phase. We emphasize that
the observation of a common strong phase $\delta$ is based on very
firm grounds within the frame work of the Standard Model (SM). It
relies essentially, only on isospin and the operator structures
contributing within the standard model. Nevertheless, an
experimentally verifiable consequence of this hypothesis would be the
vanishing of direct CP-violating asymmetry for the mode $A^{+0}\equiv
A(K^0(\pi^+\pi^0)_{\sss even})$.

Using the amplitudes $A^{+-}$, $\overline{A}^{+-}$, $A^{00}$ and
$\overline{A}^{00}$ one can construct a maximum of seven independent
observables (The amplitudes $A^{+0}$, $A^{-0}$ are not independent as
they can be obtained using isospin relations). The two triangles can
be completely defined in terms of seven observables, the three sides
of each of the triangles and a relative angle between the two
triangles.  The amplitudes under consideration involve the following
eleven variables: $a^{+-}$, $b^{+-}$, $a^{00}$, $b^{00}$, $a^{+0}$, $
b^{+0}$, $\delta^{+-}_a$, $\delta^{+-}_b$, $\delta^{00}_a$,
$\delta^{00}_b$, and $\gamma$. These variables are connected by two
isospin relations (see Eq.~(\ref{neutral-triangle}) and the
corresponding relation for the conjugate process), which results in
four constraints, reducing the number of independent variables to
seven, as we will illustrate below. Hence, all variables including
$\gamma$, can be determined purely in terms of observables.

In order to determine $\gamma$, we express all the amplitudes and
strong phases, in terms of observables and $\gamma$. The variables,
$a^{ij}$ and $b^{ij}$ may be solved as a function of $\gamma$ and
other observables as follows:
\begin{eqnarray}
 \label{eq:ab} 
|a^{+-}|^2 &=& \frac{B^{+-}}{2 \sin^2\gamma}\,
 \Bigg(1-y^{+-}\cos(2\theta^{+-})\Bigg)\\
|b^{+-}|^2 &=& \frac{B^{+-}}{2 \sin^2\gamma}\,
 \Bigg(1-y^{+-}\cos(2\theta^{+-}-2\gamma)\Bigg)\\
|a^{+0}|^2 &=& \frac{B^{+0}}{2 \sin^2\gamma}\,
 \Bigg(1-\cos(2{\tilde\gamma})\Bigg)\\
|b^{+0}|^2 &=& \frac{B^{+0}}{2 \sin^2\gamma}\,
 \Bigg(1-\cos(2{\tilde\gamma}-2\gamma)\Bigg)\\
|a^{00}|^2 &=& \frac{B^{00}}{2 \sin^2\gamma}\,
 \Bigg(1-y^{00}\cos(2\theta^{00})\Bigg)\\
|b^{00}|^2 &=& \frac{B^{00}}{2 \sin^2\gamma}\,
 \Bigg(1-y^{00}\cos(2\theta^{00}-2\gamma)\Bigg)
\label{amps}
\end{eqnarray}
where $2\theta^{00}$ defined in analogy to $2\theta^{+-}$, is the
angle between $A^{00}$ and $\overline{A}^{00}$, and $B^{+-}$, $B^{00}$
and $B^{+0}$ are given by,
\begin{equation}
B^{+-}=\frac{|\overline{A}^{+-}|^2+|A^{+-}|^2}{2},~~
B^{00}=\frac{|\overline{A}^{00}|^2+|A^{00}|^2}{2},~~
B^{+0}=\frac{|A^{-0}|^2+|A^{+0}|^2}{2}~.
\end{equation}
The angle $2\theta^{00}$ need not be measured but can
be determined from geometry of the two triangles. It is given by
\begin{equation}
 \label{eq:gamma00}
\cos\Big(2\theta^{00}-2{\tilde\gamma}\Big)=\frac{B^{00}-B^{+-} +
  |A^{+-}| |\overline{A}^{+-}|\cos(2\theta^{+-}-2{\tilde\gamma})}
        {|A^{00}| |\overline{A}^{00}|}
\label{theta00}
\end{equation}


We define $\delta^{+-}=\delta_b^{+-}-\delta_a^{+-} $ and
$\delta^{00}=\delta_b^{00}-\delta_a^{00}$, which are conveniently
expressed in terms of $\gamma$ and observables as:
\begin{eqnarray}
\tan\delta^{+-}&=&\frac{a_{\sss \rm dir}^{+-}\tan\gamma}{1-y^{+-}
[\cos2\theta^{+-}-\sin2\theta^{+-}\tan\gamma ]}\\
\tan\delta^{00}&=&\frac{a_{\sss \rm dir}^{00}\tan\gamma}{1-y^{00}
[\cos2\theta^{00}-\sin2\theta^{00}\tan\gamma ]}
\end{eqnarray}
Our task now is to express the strong phases $\delta_a^{+-}$ and
$\delta_a^{00}$ in terms of $\gamma$ and observables, just as we have
done for the other variables. One finally intends to solve for
$\gamma$, only in terms of observables.

The isospin triangle relation given by
Eq.~(\ref{neutral-triangle}) and the similar relation for the
conjugate process may be expressed as:
\begin{equation}
 (a^{+-} e^{i\delta^{+-}_a}+ a^{00} e^{i\delta^{00}_a})e^{\pm i\gamma}
      + ( b^{+-} e^{i\delta_b^{+-}}+ b^{00} e^{i\delta^{00}_b})=
      (a^{+0} e^{\pm i\gamma} + b^{+0})~.
\label{iso-triangles}
\end{equation}
Using Eq.~(\ref{iso-triangles}) one can derive the `four' equations:
\begin{eqnarray}
a^{+-}\cos(\delta_a^{+-} \pm\gamma)+a^{00}
\cos(\delta_a^{00}\pm\gamma)&+&
 b^{+-}\cos\delta_b^{+-}
   +b^{00} \cos\delta_b^{00}\nn\\ &=&
   a^{+0}\cos\gamma+b^{+0} \label{a}\\
a^{+-}\sin(\delta_a^{+-} \pm\gamma)+a^{00}
\sin(\delta_a^{00}\pm\gamma)&+&
 b^{+-}\sin\delta_b^{+-}
   +b^{00} \sin\delta_b^{00}\nn\\ &=& \pm
   a^{+0}\sin\gamma~.\label{b} 
\end{eqnarray}
Eqns.~(\ref{a}) and (\ref{b}) may be recast as follows:
\begin{eqnarray}
a^{+-}\sin\delta_a^{+-}+a^{00}\sin\delta_a^{00}&=&0  \label{1}\\
a^{+-}\cos\delta_a^{+-}+a^{00}\cos\delta_a^{00}&=& a^{+0}  \label{2}\\
b^{+-}\cos\delta_b^{+-}+b^{00}\cos\delta_b^{00}&=& b^{+0} \label{3}\\
b^{+-}\sin\delta_b^{+-}+b^{00}\sin\delta_b^{00}&=& 0 \label{4}
\end{eqnarray}
Eq.~(\ref{1}) and (\ref{2}) may be used to used to solve for
$\cos\delta_a^{+-}$ and $\cos\delta_a^{00}$:
\begin{eqnarray}
\cos\delta_a^{+-}&=&\frac{|a^{+0}|^2+|a^{+-}|^2-|a^{00}|^2}{2|a^{+0}||a^{+-}|}\\
\cos\delta_a^{00}&=&\frac{|a^{+0}|^2+|a^{00}|^2-|a^{+-}|^2}{2|a^{+0}||a^{00}|}~.
\end{eqnarray}
Squaring and adding Eqns.~(\ref{3}) and (\ref{4}) we get,
\begin{equation}
|b^{+-}|^2+|b^{00}|^2 + 2
 b^{+-}b^{00}\cos(\delta_b^{+-}-\delta_b^{00}) = | b^{+0}|^2  
\label{gamma-eq}
\end{equation}
Now $\delta_b^{+-}=\delta^{+-}+\delta_a^{+-}$ and
$\delta_b^{00}=\delta^{00}+\delta_a^{00}$. Hence, Eq.~(\ref{gamma-eq})
is expressed completely in terms of observables and $\gamma$. $\gamma$
can thus be determined cleanly, in terms of observables.

The CKM phase $\gamma$ can be determined simultaneously for several
regions of the Dalitz plot. The ambiguities in the solution of
$\gamma$ may thereby be removed.  Having measured $\gamma$
one can use Eq.~(\ref{eq:xi-gamma-rel}) to estimate the value of
$\delta_{EW}$ in terms of observables. We can thus verify our
understanding of electroweak penguin contributions.

One may ask if it is possible to determine $\gamma$ using $B\to
K\pi\pi$ without resorting to the Neubert-Rosner hypothesis. If one
includes in the analysis $B\to K_s(\pi^+\pi^-)_{\rm \sss odd}$ with
the two pions in an isospin odd state, one adds four new variables
corresponding to the amplitudes and strong phases of the two parts
with different weak phases. However, one can at best obtain four new
independent observables. Three of which, arise from time dependent
measurement for this mode, and one results from the interference
between states with pions in isospin even and isospin odd. We hence
conclude that it is not possible to determine $\gamma$ without
at-least one theoretical observation, even if one uses all the
information possible from $B\to K \pi\pi$ decays.

Current experimental data \cite{CLEO,BELLE} indicates that at least two
sides of the triangles in Fig.~\ref{fig1} are readily measurable.
While $B^+\to K_s \pi^+\pi^0$ has not yet been observed, the mode
$B^0\to K^+ \pi^-\pi^0$ has been seen. Hence, using
Eq.~(\ref{iso-rel}) the isospin triangle in
Eq.~(\ref{neutral-triangle}) can still be constructed. In fact, in
future, data from both $B^+\to K_s \pi^+\pi^0$ and $B^0\to K^+
\pi^-\pi^0$ modes could be combined to improve statistics.

To conclude, the weak phase $\gamma$ can be measured using a time
dependent asymmetry measurement in the three body decay, $B\to
K\pi\pi$. A detailed study of the Dalitz plot can be used to extract
the $\pi\pi$ even isospin states. These states obey certain isospin
relations which allow us to not only obtain $\gamma$, but also
determine the size of the electroweak penguin contribution. In
contrast to methods of determination of $\gamma$ using the two body
decay modes $B\to K\pi$, this technique does not require any
theoretical assumptions like SU(3) or neglect of any contributions to
the decay amplitudes. By studying different regions of the Dalitz plot
it is possible to reduce the ambiguity in the value of $\gamma$.

\section*{\bf Acknowledgments}
This work was partly supported by DOE Grant DE-FG03-96ER40969. The
work of N.S. was supported by the young scientist award of the
Department of Science and Technology, India.


\begin{thebibliography}{References:}


  
\bibitem{CarterSanda}A. B. Carter and A. I. Sanda \prl{45}{80}{952};
  A. B. Carter and A.I. Sanda, \prd{23}{81}{1567}; I. Bigi and A. I.
  Sanda, Nucl. Phys. {\bf B193}, 85 (1981).
  
\bibitem{betameas} B.~Aubert {\it et al.}  [BABAR Collaboration],
arXiv:hep-ex/0203007; M.~Hazumi {\it et al.}  [BELLE Collaboration],
talk given the at ``Flavor Physics and $CP$ Violation (FPCP)'', May
16-18, 2002, Philadelphia, (U.S.A.);
R. Patterson,  Summary talk given the at 
``Flavor Physics and $CP$ Violation (FPCP)'', May 16-18, 2002,
Philadelphia, (U.S.A.).

  
\bibitem{Hoecker}A. Hoecker, Talk given at 
``Flavor Physics and $CP$ Violation (FPCP)'', May 16-18, 2002,
Philadelphia, (U.S.A.).


\bibitem{GL}M.  Gronau and D. London, \prl{65}{90}{3381}.  

\bibitem{gamma-tech}M.Gronau and D.~London,
  { Phys. Lett.} {\bf B253}, 483 (1991); M.Gronau and D.~Wyler,
  { Phys. Lett.} {\bf B265}, 172 (1991); I.~Dunietz, { Phys.
  Lett.} {\bf B270}, 75 (1991); D.~Atwood, I.  Dunietz and A. Soni,
  { Phys. Rev. Lett.}  {\bf 78}, 3217 (1997);
 N.~Sinha and R.
  Sinha, \prl{80}{98}{3706}; K.~Agashe and N.~G.~Deshpande,
 { Phys. Lett.} {\bf  B451}, 215 (1999).

\bibitem{SU3} 
M.~Gronau,
arXiv:hep-ph/0104050, Invited talk at the 4th International Conference
on B Physics and CP Violation (BCP4), Ise-Shima, Japan, Febuary
18-23,2001 and references therein.

\bibitem{CLEO}
E.~Eckhart {\it et al.}  [CLEO Collaboration],
arXiv:hep-ex/0206024.
\bibitem{BELLE}
A.~Garmash, [on behalf of Belle Collaboration], Talk given at 
``Flavor Physics and $CP$ Violation (FPCP)'', May 16-18, 2002,
Philadelphia, (U.S.A.).



\bibitem{LNQS}
H.~J.~Lipkin, Y.~Nir, H.~R.~Quinn and A.~Snyder,
Phys.\ Rev.\ D {\bf 44}, 1454 (1991).

\bibitem{DH}
N.~G.~Deshpande and X.~G.~He,
\prl {\bf 74}, 26 (1995)
[Erratum-ibid.\  {\bf 74}, 4099 (1995)]

\bibitem{CKM}N.  Cabibbo,{\it Phys. Rev.  Lett.} {\bf 10}, 531 (1963);
  M. Koba\-yashi and T.~Maskawa, {\it Prog. Theor. Phys.} {\bf 49},
  652 (1973); L. Wolfenstein, \prl{51}{83}{1945}.

\bibitem{NR}
M.~Neubert and J.~L.~Rosner,
Phys.\ Lett.\ B {\bf 441}, 403 (1998).


\bibitem{GPY}
M.~Gronau, D.~Pirjol and T.~M.~Yan,
Phys.\ Rev.\ D {\bf 60}, 034021 (1999); M.~Gronau and D.~Pirjol,
Phys.\ Rev.\ D {\bf 62}, 077301 (2000).

\end{thebibliography}
\end{document}